\documentclass[final,5p,times]{elsarticle}
\usepackage{amsmath,amssymb,bm,graphicx,multicol,color,soul,dsfont,multirow}
\usepackage[colorlinks=true, pdfstartview=FitV, linkcolor=blue, citecolor= blue, urlcolor=blue]{hyperref}
\biboptions{sort&compress}
\allowdisplaybreaks
\journal{Physics Letters B}

\begin{document}
\begin{frontmatter}
\title{The phenomenology of the exotic hybrid nonet with \texorpdfstring{$\pi_1(1600)$}{} and \texorpdfstring{$\eta_1(1855)$}{}}
\author[ujk]{Vanamali Shastry\corref{1}}
\ead{vanamalishastry@gmail.com}

\author[jlu,gsi]{Christian S. Fischer}
\ead{christian.fischer@theo.physik.uni-giessen.de}

\author[ujk,guf]{Francesco Giacosa}
\ead{fgiacosa@ujk.edu.pl}

\cortext[1]{Corresponding author}
\address[ujk]{Institute of Physics, Jan Kochanowski University, ul.  Uniwersytecka  7,  P-25-406  Kielce,  Poland}
\address[jlu]{Institut f\"ur Theoretische Physik, Justus-Liebig Universit\"at Gie\ss en, 35392 Gie\ss en, Germany}
\address[gsi]{Helmholtz Forschungsakademie Hessen f\"ur FAIR (HFHF), GSI Helmholtzzentrum f\"ur Schwerionenforschung, Campus Gie\ss en, 35392 Gie\ss en, Germany}
\address[guf]{Institute for Theoretical Physics, Johann Wolfgang Goethe - University, Max von Laue--Str. 1 D-60438 Frankfurt, Germany}

\begin{abstract}
We study the decays of the $J^{PC}=1^{-+}$ hybrid nonet using a Lagrangian invariant under the flavor symmetry, parity reversal, and charge conjugation. We use the available experimental data, the lattice predictions, and the flavor constraints to evaluate the coupling strengths of the $\pi_1(1600)$ to various
two-body mesonic states. Using these coupling constants, we estimate the partial widths of the two-body decays of the hybrid pion, kaon and the isoscalars. We find that the hybrid kaon can be nearly as broad as the $\pi_1(1600)$. Quite remarkably, we find also that the light isoscalar must be significantly narrow while the width of the heavy isoscalar can be matched to the recently observed $\eta_1(1855)$.
\end{abstract}
\begin{keyword}
hybrid mesons, chiral lagrangian, meson decays.
\end{keyword}

\end{frontmatter}

\section{Introduction}
According to the non-relativistic quark model, conventional mesons ($\bar{q}q$ states) can only have specific values 
for the spin ($J$), parity ($P$), and charge conjugation ($C$) quantum numbers. These values are determined by the 
spin and angular momentum of the constituent quarks, which are conserved separately in a non-relativistic framework.  
States with unconventional quantum numbers, {\it e.g., $J^{PC} = 0^{--}$, $0^{+-}$, $1^{-+}$, $2^{+-},\ldots,$} are 
termed exotic. These exotic quantum numbers can arise due to various mechanisms like relativity, addition of non-$\bar{q}q$ 
degrees of freedom (like gluons) or the formation of multiquark bound states. All these may lead to a rich ``exotic" spectrum 
of QCD involving, amongst others, glueballs, four-quark states and quark-gluon hybrids. 

Although a number of candidates for exotic states have been observed experimentally, the identification of whole multiplets
is far from complete. There is ample evidence for the presence of four- and five-quark states in the heavy quark sector of QCD, 
see {\it e.g} Ref.~\cite{Brambilla:2019esw} for a review. In the light quark sector two potential hybrid states,
the isovector $\pi_1(1600)$ \cite{Zyla:2020ssz} and the very recently observed isoscalar $\eta_1(1855)$ \cite{Ablikim:2022zze},
have been identified as states with quantum numbers $J^{PC} = 1^{-+}$. The observation of the hybrid isoscalar has 
piqued the interests of the community \cite{Chen:2022qpd,Qiu:2022ktc,Dong:2022cuw}, since it may provide important 
guidance for high-quality predictions of masses and decay widths of the missing members of the $J^{PC} = 1^{-+}$-nonet.  
The members of (light) hybrid nonet are the subject matter of the present study.

First observed in 1998 by the E852 collaboration in the $\pi\text{-}p$ scattering process \cite{E852:1998mbq}, the mass
and width of the $\pi_1(1600)$ have been determined as $1661^{+15}_{-11}$ MeV and $240\pm 50$ MeV 
\cite{Zyla:2020ssz}. This state has been observed to decay into $b_1\pi$, $\rho\pi$, $\eta'\pi$, and $f_1\pi$ \cite{Zyla:2020ssz} and it can be identified with the light hybrid isovector predicted by models \cite{Meyer:2015eta}; its partners
in the $1^{-+}$ nonet are expected to have masses less than $2$ GeV. An additional predicted decay channel, not yet 
observed by experiment is the $\eta\pi$ channel. Interestingly, the experiments have reported a second (lighter) resonance 
with hybrid quantum numbers $1^{-+}$, called the $\pi_1(1400)$, that decays into $\eta\pi$ \cite{E862:2006cfp,OBELIX:2004oio,CrystalBarrel:1999reg,CrystalBarrel:1998cfz}.
The fact that there are two hybrid isovector states with the same quantum numbers but mutually exclusive decay channels 
suggests that these two states must be the same, and the difference in the observed masses must be due to interference 
of background processes \cite{Bass:2001zs}. This hypothesis has been extensively studied and corroborated by Ref.~\cite{JPAC:2018zyd}. 
Besides the $\pi_1$ states, the search for additional hybrid states represents an ongoing experimental effort with contributions from various collaborations 
\cite{COMPASS:2021ogp,Hamdi:2019dbr,BESIII:2022qzu}.
\begin{table*}[t]
        \centering
        \renewcommand{\arraystretch}{1.5}
        \begin{tabular}{|c|c|c|c|c|c|c|c|}
        \hline
             Meson & Mass (GeV) & Meson & Mass (GeV) &Meson & Mass (GeV) & States & Mixing angle\\\hline
             $\pi$ & 0.135 & $K$ & 0.494 & $\eta$ & $0.548$ & $\eta-\eta^\prime$ & $-44.5^\circ$ \cite{Amelino-Camelia:2010cem}\\
             $\eta^\prime$ & 0.958 & $\rho$ & 0.775 & $\omega$ & $0.782$ & $\phi-\omega$ & $-3^\circ$\\
             $K^*$ & 0.892 & $\phi$ & 1.020 & $a_1$ & $1.23$ & $f_1-f_1^\prime$ & $24^\circ$ \cite{LHCb:2013ged}\\
             $K_1(1270)$ & 1.253 & $f_1$ & 1.285 & $f_1^\prime$ & $1.426$ & $h_1-h_1^\prime$ & $25^\circ$ \cite{Shastry:2021asu}\\
             $b_1$ & $1.23$ & $K_1(1400)$ & 1.403 & $h_1(1170)$ & $1.17$ & $K_1^A - K_1^B$ & $56^\circ$ \cite{Divotgey:2013jba,Cheng:2011pb}\\\hline
        \end{tabular}
        \caption{The values of the masses and mixing angles of the decay products used in the fit. All values taken from the PDG \cite{Zyla:2020ssz} unless otherwise noted.}
        \label{prodtab}
    \end{table*}
    \begin{table*}[ht]
    \centering
    {\renewcommand{\arraystretch}{1.5}
    \begin{tabular}{|c|c|c|}
    \hline
        Parameter & \multicolumn{2}{c|}{Value} \\\cline{2-3}
        & Set$-1$ ($D/S>0$) & Set$-2$ ($D/S<0$)\\\hline
        $m_{\pi_1}$ & $1.663\pm 0.01$ GeV & $1.662\pm 0.01$\\
        $g_{\scriptscriptstyle b_1\pi}^c$ & $88\pm 23$ GeV & $-(119 \pm 22)$\\
        $g_{\scriptscriptstyle b_1\pi}^d $ & $-(23.3\pm 5.60)$ GeV$^{-1}$ & $26.7\pm 5.3$\\
        $g_{\scriptscriptstyle \rho\pi}$ & $0.35\pm 0.05)$ GeV & $0.35\pm 0.05$\\
        $g_{\scriptscriptstyle f_1\pi}$ & $8.02\pm 0.83$ GeV & $8.12\pm 0.83$\\
        $g{\scriptscriptstyle \rho\omega}$ & $-(0.37 \pm 0.07)$ & $-(0.38\pm 0.07)$\\
        $g_{\scriptscriptstyle \eta\pi}$ & $4.91\pm 0.56$ & $4.94\pm 0.55$\\\hline
        $\chi^2/$d.o.f & $0.35$ & $0.28$ \\\hline
    \end{tabular}}
    \caption{The values of the mass of $\pi_1(1600)$ and the coupling constants along with the uncertainties when the $D/S$-ratio for the $b_1\pi$ decay channel is positive (second column), and negative (third column).}
    \label{parfit}
\end{table*}

On the theoretical side, continuum methods such as quark models \cite{Isgur:1985vy,Page:1998gz,Burns:2006wz,Close:1994hc,Swanson:1997wy}, 
bag models \cite{Chanowitz:1982qj}, QCD sum rules \cite{Chen:2010ic,Huang:2010dc}, functional methods via 
Dyson-Schwinger, Bethe-Salpeter and Faddeev equations \cite{Burden:2002ps,Krassnigg:2009zh,Qin:2011xq,Hilger:2015hka,Williams:2015cvx,Xu:2018cor}, 
light-front quantization models \cite{Lan:2021wok} and coupled-channel $K$-matrix analysis \cite{Kopf:2020yoa}
have been used to study the properties of hybrid mesons. These are complemented by various lattice studies 
exploring the exotic side of the QCD spectrum resulting in various qualitative predictions
\cite{Lacock:1996vy,Lacock:1996ny,MILC:1997usn,McNeile:2006bz,Dudek:2010wm,Dudek:2011bn,Dudek:2011tt,Dudek:2013yja}. A recent comprehensive study 
of the hybrid (isovector) mesons on the lattice reported the mass of the $\pi_1(1600)$ to be $\sim 1.564$ GeV \cite{Woss:2020ayi} and also the possible ranges for the partial widths of the decays of the $\pi_1(1600)$. 

In the present work, we combine in a unique fit, experimental data \cite{Zyla:2020ssz}, lattice QCD results \cite{Woss:2020ayi}, as well as constraints from flavour symmetry in order to constraint the mass and the decay widths of the resonance $\pi_1(1600)$. [As a means, we shall construct a Lagrangian invariant under flavor symmetry, parity reversal and charge conjugation, whose coupling constants are fixed by the aforementioned fit.] 
As a result, we are able to estimate the mass and the decays of the $\pi_1(1600)$ to a better accuracy than the present experimental and lattice results.\par

Moreover, flavor symmetry implies that hybrid mesons also appear in nonets, just as it happens for the regular quark - antiquark states. Within our approach, we are able to predict the decay properties of the remaining members of the nonet {\it viz.,} the kaonic and the isoscalar ones. In particular, we find that the light isoscalar 
state ($\eta_1^L$) is quite narrow, but the heavy isoscalar state ($\eta_1^H$) can be as broad as $\sim 200$ MeV. The latter, in fact, is consistent with the recently observed $\eta_1(1855)$ which has a width of $188\pm 18^{+3}_{-8}$ MeV \cite{Ablikim:2022zze}. This is quite interesting, since the existence of these states could be verified in ongoing and future experiments.\par

The paper is divided into the following sections: in Sec. \ref{chisym} we briefly discuss the Lagrangian for $\pi_1$ and in Sec. \ref{secfit} we perform the fit. In Sec. \ref{secpred} we present the our results for other members of the nonet and discuss their possible implications; in Sec. \ref{secsum} we summarize the paper.\par

\begin{table*}[ht]
    \centering
    {\renewcommand{\arraystretch}{1.5}
 \begin{tabular}{|c|c|c|c|}
    \hline
        Channel & Width (MeV) & Channel & Width (MeV)\\\hline
        $\Gamma_{b_1\pi}$ & $220\pm 34$ & $\Gamma_{f_1\pi}$ & $16.2\pm 3.1$\\
        $\Gamma_{\rho\pi}$ & $7.1\pm 1.8$ & $\Gamma_{f_1^\prime\pi}$ & $0.83\pm 0.16$\\
        $\Gamma_{K^* K}$ & $1.2\pm 0.3$ & $\Gamma_{\eta\pi}$ & $0.37\pm 0.08$\\
        $\Gamma_{\rho\omega}$ & $0.08\pm 0.03$ & $\Gamma_{\eta^\prime\pi}$ & $4.6\pm 1.0$ \\\hline
        & & $\Gamma_\text{tot}$ & $250\pm 34$ \\\hline
    \end{tabular}}
    \caption{The partial widths and branching ratios of various decay channels and the total width  (parameter Set-1; see text for discussion).}
    \label{decwid}
\end{table*}

\section{The Lagrangian for the \texorpdfstring{$\pi_1$}{} state}\label{chisym}
In order to describe the decays of the state $\pi_1(1600)$, we write down a simple Lagrangian containing the relevant interaction terms and respecting invariance under flavor symmetry, parity reversal ($P$) and charge conjugation ($C$): 
\begin{align}
    \mathcal{L}^\pi_{hyb} &= g_{b_1\pi}^c \langle \pi_{1,\mu}b_1^\mu\pi\rangle+ g_{b_1\pi}^d \langle \pi_{1,\mu\nu}b_1^{\mu\nu}\pi\rangle\nonumber\\
    & + g_{f_1\pi} \langle \pi_{1,\mu}f_{1,N}^{\mu\nu}\partial_\nu\pi + \pi_{1,\mu}f_{1,S}^{\mu\nu}\partial_\nu\pi\rangle\nonumber\\
    &  + g_{\eta\pi} \langle\pi_{1,\mu} (\eta_N \partial^\mu\pi + \eta_S \partial^\mu\pi)\rangle +g_{\rho\pi}\langle \tilde{\pi}_{1,\mu\nu} \rho^{\mu\nu}\pi \rangle\nonumber\\
    & + g_{\rho\omega} \langle \pi_{1,\mu}(\rho^{\mu\nu}\omega_\nu+\omega^{\mu\nu}\rho_\nu)\rangle.\label{lagF}
\end{align}
In the above Lagrangian, the isospin factors have been represented as $\langle \cdots \rangle$, and the subscripts $N$ and $S$ represent non-strange and strange flavor states respectively. The first two terms describe the decay of the $\pi_1$ to $b_1\pi$, but the latter is a higher order term that includes derivatives. This has been included to take care of the large ratio of the partial wave amplitudes in the decay \cite{Zyla:2020ssz,Shastry:2021asu}.
The term in the second line of Eq. \ref{lagF} gives rise to the decay of the $\pi_1$ to the axial-vector isoscalars. As we demonstrate later, when extended to the entire nonet, this term describes some of the dominant decays of the kaonic hybrid and the isoscalars. The term describing the $\eta^{(\prime)}\pi$ decay channels is peculiar, since it arises entirely due to the axial anomaly \cite{Eshraim:2020ucw}. The last line of Eq. \ref{lagF} describes the decay of the hybrid into two vector states. These channels, however, turns out to be highly suppressed, similarly to the $\eta\pi$ channel.\par

In summary, the Lagrangian above can be seen as a tool to summarize the available decay channels, allowing us to write down the corresponding decay widths in each case. The coupling constants, that shall be determined by a fit to data and lattice, contain the (non-trivial) link between the hybrid state $\pi_1$ and the ordinary mesons. Eventual form factors, and other non-perturbative effects are absorbed into the values of the coupling constants.\par

From Eq. \ref{lagF}, we get the following expressions for the decay widths:
\begin{align}
    \Gamma_{b_1\pi}&=\frac{1}{2}\frac{k_{b_1}}{24\pi m_{\pi_1}^2}\left(\frac{1}{m_{b_1}^2}\left(E_{b_1} g_{b_1\pi}^c+2 g_{b_1\pi}^d m_{b_1}^2 m_{\pi_1}\right)^2\right.\nonumber\\
    & \left. +2 (2 E_{b_1} g_{b_1\pi}^2 m_{\pi_1}+g_{b_1\pi}^c)^2\right)\\
    \frac{G_2}{G_0}&= \sqrt{2} \frac{g_{b_1\pi}^c (-E_{b_1} + m_{b_1}) + 2 g_{b_1\pi}^d m_{\pi_1} (-m_{b_1}^2 + E_{b_1} m_{b_1})}{g_{b_1\pi}^c (E_{b_1} + 2 m_{b_1}) + 2 g_{b_1\pi}^d m_{\pi_1} (m_{b_1}^2 + 2 E_{b_1} m_{b_1})}\label{dsrat}\\
   \Gamma_{\rho\pi}&=g_{\rho\pi}^2\frac{ k_\rho^3}{6 \pi}\label{Grp}\\
   \Gamma_{K^*K}&=g_{\rho\pi}^2\frac{ k_{K^*}^3}{12 \pi}\label{Gksk}\\
   \Gamma_{f_1\pi} &=\frac{1}{4}g_{f_1\pi}^2 \cos^2\theta_a \frac{k_{{f_1}}}{24\pi m_{\pi_1}^2}\left(m_{f_1}^2 (2 m_{f_1}^2 + m_{\pi_1}^2 - 6 m_{\pi_1}E_{f_1}) \right.\nonumber\\
    & \left. + E_{f_1}^2 (2 m_{\pi_1}^2 + m_{f_1}^2)\right)\label{Gf1p}\\
 \Gamma_{f_1^\prime\pi} &=\frac{1}{4}g_{f_1\pi}^2 \sin^2\theta_a \frac{k_{f_1^\prime}}{24\pi m_{\pi_1}^2}\left(m_{f_1^\prime}^2 (2 m_{f_1^\prime}^2 + m_{\pi_1}^2 - 6 m_{\pi_1}E_{f_1^\prime})  \right.\nonumber\\
    & \left. + E_{f_1^\prime}^2 (2 m_{\pi_1}^2 + m_{f_1^\prime}^2)\right)\label{Gf1pp}\\
   \Gamma_{\eta\pi} &= \frac{1}{4}\left(\sqrt{\frac{2}{3}}\cos\theta_p + \frac{1}{\sqrt{3}}\sin\theta_p \right)^2 g_{\eta}^2 \frac{k_{\eta}^3}{24\pi m_{\pi_1}^2}\label{Gnp}\\
   \Gamma_{\eta^\prime\pi} &=  \frac{1}{4}\left(\sqrt{\frac{2}{3}}\sin\theta_p - \frac{1}{\sqrt{3}}\cos\theta_p \right)^2 g_\eta^2 \frac{k_{\eta^\prime}^3}{24\pi m_{\pi_1}^2}\label{Gn1p}\\
   \Gamma_{\rho\omega} &= \frac{\cos^2\theta_v}{4}g_{\rho\omega}^2\frac{k_\rho^3}{24\pi} \left(\frac{\left(-8 E_{\rho} m_{\pi_1}+4 m_{\rho}^2+4 E_{\rho}^2+3
   m_{\pi_1}^2\right)}{m_{\rho}^2 m_{\omega}^2}\right),
\end{align}
where $m_{\pi_1}$ is the mass of the parent, $G_2$ and $G_0$ are the amplitudes of the $\ell=0,2$ partial waves, $k_x$ and $g_x$ are the 3-momenta carried by the decay products and the coupling constants in the channel $x$ respectively. In the expressions above, the notations $f_1$ and $f_1^\prime$ represent $f_1(1285)$ and $f_1^\prime(1420)$. Further, $\theta_p$ is the $\eta$-$\eta^\prime$ mixing angle, $\theta_v$ is the $\omega-\phi$ mixing angle, and $\theta_a$ is the $f_1$-$f_1^\prime$ mixing angle in the strange-nonstrange basis. The values of the masses of the final states and, wherever applicable, the mixing angles are listed in Table \ref{prodtab}.

\begin{table*}[ht]
\centering
    {\renewcommand{\arraystretch}{1.5}\begin{tabular}{|c|c|c|c|c|c|c|}
        \hline
        &  $m_{K_1}$(GeV) & $m_{\eta_1^L}$(GeV) & $m_{\eta_1^H}$(GeV)& $\theta_h$ & $\delta_S^{hyb}$ (GeV$^2$)\\\hline
        Scenario-1 & $1.707$ & $1.542$ & $1.855$ & $36.7^\circ$ & $0.151$\\
        Scenario-2 & $1.761$ & $1.661$ & $1.855$ & $0^\circ$ & $0.341$\\
        Scenario-3 & $1.754$ & $1.646$ & $1.855$ & $15^\circ$ & $0.317$\\\hline
    \end{tabular}}
    \caption{The masses of the kaons and the isoscalars, and isoscalar mixing angle for the three scenarios discussed in the text.}\label{parscen}
\end{table*}
\begin{table*}[ht]
        \centering
        {\renewcommand{\arraystretch}{1.5}\begin{tabular}{|c|c|c|c|c|c|c|c|}
             \hline
             State &\multicolumn{2}{c|}{Scenario-1}&\multicolumn{2}{c|}{Scenario-2}&\multicolumn{2}{c|}{Scenario-3} & Parameter\\\cline{2-7}
             & M (MeV) & $\Gamma$ (MeV) & M (MeV) & $\Gamma$ (MeV) & M (MeV) & $\Gamma$ (MeV) & Set\\\hline
             \multirow{2}{*}{$K_1^{hyb}$} & \multirow{2}{*}{$1706$} & $140\pm 45$ &  & $\bm{312\pm 97}$ & \multirow{2}{*}{$1754$} & $286\pm 88$ & 1\\
             & & $73\pm 29$ & \raisebox{2.2ex}[2.2ex]{$\bm{1761}$} & $\bm{170\pm 65}$ & & $155\pm 59$ & 2\\\hline
              \multirow{2}{*}{$\eta_1^L$} & \multirow{2}{*}{$1543$} & $21\pm 4$ & & $\bm{81\pm 15}$ & \multirow{2}{*}{$1646$} & $69\pm 13$ & 1\\
             & & $22\pm 4$ & \raisebox{2.2ex}[2.2ex]{$\bm{1661}$} & $\bm{83\pm 16}$ & & $71\pm 13$ & 2\\\hline
             \multirow{2}{*}{$\eta_1^H$} & \multirow{2}{*}{$1855$} & $607\pm 159$ &  & $\bm{259\pm 92}$ & \multirow{2}{*}{$1855$} & $411\pm 130$ & 1\\
             & & $249\pm 80$ & \raisebox{2.2ex}[2.2ex]{$\bm{1855}$} & $\bm{157\pm 68}$ & & $192\pm 80$ & 2\\\hline
        \end{tabular}}
        \caption{The masses and widths of the kaons and the isoscalars in the three scenarios discussed in the text. The most probable scenario is highlighted in bold.}
        \label{tabsum}
    \end{table*}

\section{Combined fit for \texorpdfstring{$\pi_1(1600)$}{}}\label{secfit}
The available experimental data on the properties of $\pi_1(1600)$, used in our fit, are:
\begin{enumerate}
    \item The mass of the hybrid: $m_{\pi_1}=1661^{+15}_{-11}$ MeV  \cite{Zyla:2020ssz}. [For the different experimental results considered by the PDG to arrive at this value, see Refs. \cite{COMPASS:2018uzl,E852:2004rfa,E852:2004gpn,E852:2001ikk}]. We take the uncertainty in the mass to be $15$ MeV.
    \item The decay width: $\Gamma_\text{tot}=240\pm50$ MeV \cite{Zyla:2020ssz}.
    \item The ratio of the branching ratios of the $b_1\pi$ channel in the $D$-wave and $S$-wave: $\dfrac{BR(\pi_1\to b_1\pi)_D}{BR(\pi_1\to b_1\pi)_S}=0.3\pm0.1$ \cite{Zyla:2020ssz,Baker:2003jh}. This ratio is equal to the square-root of the ratio of the corresponding partial wave amplitudes (PWAs). Hence, the $D/S$-ratio for the $b_1\pi$ channel is $$\sqrt{\dfrac{BR(\pi_1\to b_1\pi)_D}{BR(\pi_1\to b_1\pi)_S}}=\pm(0.55\pm0.165).$$ The sign of the $D/S$-ratio is, unfortunately, unknown. Thus, we perform two fits -- one for each sign of this ratio. From Eq. \ref{dsrat} we see that the $D/S$-ratio fixes the magnitude of the ratio of the the coupling constants $g_{b_1\pi}^d/g_{b_1\pi}^c$. It should be noted, however, the relative sign of the coupling constants does not reflect that of the partial wave amplitudes.
    \item The ratio of the partial widths of the $f_1\pi$ channel to that of the $\eta^\prime\pi$ channel is $\dfrac{\Gamma_{f_1\pi}}{\Gamma_{\eta^\prime\pi}}=3.8\pm0.78$ \cite{Zyla:2020ssz,E852:2004gpn}.
\end{enumerate}
The ranges of the partial widths of the decay channels were estimated using lattice methods in Ref. \cite{Woss:2020ayi}. We use the midpoints of their values and estimate the uncertainty to be $50\%$. For example, the lattice result for the partial width of the $b_1\pi$ decay channel is $139$-$529$ MeV, then we use the value $334\pm 167$ MeV as an input in the fit. The following are the lattice estimates \cite{Woss:2020ayi}:
\begin{multicols}{2}
\begin{enumerate}
    \item $\Gamma_{b_1\pi}=139$-$529$ MeV,
    \item $\Gamma_{\rho\pi}=0$-$20$ MeV,
    \item $\Gamma_{K^\ast K}=0$-$2$ MeV,
    \item $\Gamma_{f_1\pi}=0$-$24$ MeV,
    \item $\Gamma_{f_1^\prime\pi}=0$-$2$ MeV,
    \item $\Gamma_{\rho\omega}\le 0.15$ MeV,
    \item $\Gamma_{\eta\pi}=0$-$1$ MeV,
    \item $\Gamma_{\eta^\prime\pi}=0$-$12$ MeV.
\end{enumerate}
\end{multicols}
Finally, we can also use the following flavor constraints:
\begin{enumerate}
    \item The $f_1$ and the $f_1^\prime$ are isoscalar doublets and arise as a results of the mixing between the corresponding strange and non-strange flavor states. Since the $\pi_1(1600)$ is an isovector, the coupling constants defining the $\pi_1 f_1^{(\prime)} \pi$ interactions differ only in the mixing angle. The decay widths also differ in the 3-momenta carried by the decay products. Thus, from Eq. \ref{Gf1p} and Eq. \ref{Gf1pp}, the ratio of the partial decay widths is $\dfrac{\Gamma_{f_1^\prime\pi}}{\Gamma_{f_1\pi}}=0.0512$, when $m_{\pi_1}=1661$ MeV.
    \item The above argument also applies to the $\rho\pi$ and $K^*K$ channels, where the difference lies in the 3-momenta and the isospin factors only. We can derive the ratio of the partial widths using Eq. \ref{Grp} and Eq. \ref{Gksk} as $\dfrac{\Gamma_{K^*K}}{\Gamma_{\rho\pi}}=0.178$. 
    \item The partial widths of the $\eta\pi$ and $\eta^\prime\pi$ channels differ in the 3-momenta, isospin factors, and the mixing angle. Using these arguments, we obtain from Eq. \ref{Gnp} and Eq. \ref{Gn1p}, $\dfrac{\Gamma_{\eta^\prime\pi}}{\Gamma_{\eta\pi}}=12.72$.
    \item We assume an arbitrary $30\%$ error in all the inputs coming from the chiral constraints.
\end{enumerate}

The values of the parameters were estimated using a $\chi^2$-fit to the available data and are listed in Table \ref{parfit}. Table \ref{decwid} lists the partial widths of the various decays of the $\pi_1(1600)$.\par

\begin{table*}[ht]
    \centering
    {\renewcommand{\arraystretch}{1.5}
    \begin{tabular}{|c|c|c||c|c|c|}
    \hline
        Channel & \multicolumn{2}{c||}{Width (MeV)}& Channel & \multicolumn{2}{c|}{Width (MeV)} \\\cline{2-3}\cline{5-6}
        & Set$-1$ & Set$-2$ & & Set$-1$ & Set$-2$\\\hline
        $\Gamma_{a_1\pi}$ & $80\pm 15$ & $82\pm 16$ & $\Gamma_{K_1(1270)K}$ & $253\pm 92$ & $151\pm 67$\\
        $\Gamma_{K^* K}$ & $0.29\pm 0.075$ & $0.29\pm0.075$ & $\Gamma_{K^*K}$ & $1.45\pm 0.37$ & $1.46\pm 0.38$ \\
        $\Gamma_{\eta^\prime\eta}$ & $0.41\pm 0.09$ & $0.41\pm0.09$ & $\Gamma_{\eta^\prime\eta}$ & $2.28\pm 0.51$ & $2.31\pm 0.51$ \\
        $\Gamma_{K_1(1270)K}$ & $0$ & $0$ & $\Gamma_{a_1\pi}$ & $0$ & $0$ \\
        $\Gamma_{\rho\rho}$ & $0.081\pm 0.028$ & $0.082\pm 0.029$ & $\Gamma_{\rho\rho}$ & $0$ & $0$ \\
        $\Gamma_{K^*K^*}$ & $0$ & $0$ & $\Gamma_{K^*K^*}$ & $0.075\pm 0.027$ & $0.077\pm 0.028$ \\
        $\Gamma_{\omega\phi}$ & $0$ & $0$ & $\Gamma_{\omega\phi}$ & $\sim 10^{-4}$ & $\sim 10^{-4}$ \\
        $\Gamma_{f_1\eta}$ & $0$ & $0$ & $\Gamma_{f_1\eta}$ & $2.15\pm 0.56$ & $2.21\pm 0.57$ \\\hline
        $\Gamma_\text{tot}$ & $81\pm 15$ & $83\pm 16$ & $\Gamma_\text{tot}$ & $259\pm 92$ & $157\pm 68$\\\hline
    \end{tabular}}
     \caption{The partial widths and branching ratios of various decay channels and the total width of the $\eta_1^L$ (left) and the $\eta_1(1855)$ (right) for $\theta_h=15^\circ$. This corresponds to the ``Scenario-3" discussed in the text.}
    \label{decwideta}
\end{table*}

It is clear that the most dominant channel for the decay of the $\pi_1(1600)$ is the $b_1(1235)\pi$ channel. As discussed in Ref. \cite{Shastry:2021asu}, derivative interactions are needed to explain the large $D/S$ ratio for the $\pi_1(1600)\to b_1(1235)\pi$ decay mentioned in the PDG \cite{Zyla:2020ssz,Baker:2003jh}. The difference in the sign of the $D/S$ ratio does not affect the results of the fit or the values of the parameters. However, the kaonic and the isoscalar hybrid states (see discussions below) show strong sensitivities to this sign. Thus, from a theoretical point of view, it would be crucial to know the phase difference between the $S$-wave and the $D$-wave in the $\pi_1(1600)\to b_1\pi$ decay. On the other hand, a knowledge of the total width of the kaons can hint at the correct sign of the $D/S$-ratio.\par

The other decays are smaller than 20 MeV, but $f_1 \pi$, $\rho \pi$, and $\eta^{\prime} \pi$ are however not negligible.  
The remaining channels {\it viz.,} $K^*K$, $f_1^\prime\pi$, and $\rho\omega$, are largely suppressed. 
We also note that  the partial width of the $\eta\pi$ decay channel is nearly one order of magnitude smaller than the  $\eta'\pi$ one. This results from the specific form of the factor dependent on the mixing angle (see, Eq. \ref{Gnp} and Eq. \ref{Gn1p}) which suppresses the contributions of the $\eta\pi$ channel by a factor of $\sim 30$ compared to that of the $\eta^\prime\pi$.

An important remark is in order: the smallness of the $\eta \pi$  channel may imply that the background effects, such as final state interactions, may influence the $\eta\pi$ channel more than the $\eta'\pi$ channel. This fact can explain the observed $\pi_1(1400)$ resonance with mass $\sim 1350$ MeV \cite{JPAC:2018zyd,Bass:2001zs} as the very same state $\pi_1(1600)$, whose peak is shifted \cite{JPAC:2018zyd}. A detailed study of this effect is left for the future. 

\section{Predictions for the other hybrid members}\label{secpred}
With the parameters determined above, we use flavor symmetry to predict the magnitude of the decays of the remaining members of the hybrid nonet  - the kaonic and the isoscalar ones. \par

\subsection{The hybrid isoscalars: \texorpdfstring{$\eta_1^L$}{} and \texorpdfstring{$\eta_1^H$}{}}
Referring to the isoscalars, the flavor states are the strange ($|\bar{s}s\rangle_h$) and the non-strange ($|\bar{n}n\rangle_h$) states. As per our model, the non-strange state is expected to have the same mass as that of the isovector. The strange state, however, is expected to get an additional mass proportional to the contributions from the strange constituent quarks (hereafter, strangeness contribution) \cite{Eshraim:2020ucw}. The masses of these states are,
\begin{align}
    m_{\eta_{1,N}}^2 &= m_{\pi_1}^2\label{metan}\\
    m_{\eta_{1,S}}^2 &= m_{\pi_1}^2 + 2\delta_S^{hyb}\label{metah}
\end{align}
where, $\delta_S^{hyb}$ is the strangeness contribution, and the subscripts $N$ and $S$ refer to non-strange and strange states, respectively. The mixing of these two states lead to the physical hybrid isoscalar mesons $\eta_{1}^L$ and $\eta_{1}^H$ (L = light, H = heavy) given by:
\begin{align}
     \begin{pmatrix}
    |\eta_1^L\rangle\\
    |\eta_1^H\rangle
    \end{pmatrix}&= \begin{pmatrix}
    \cos\theta_h & \sin\theta_h\\-\sin\theta_h & \cos\theta_h
    \end{pmatrix}
    \begin{pmatrix}
    |\bar{n}n\rangle_h\\
    |\bar{s}s\rangle_h
    \end{pmatrix}\label{etamix}
\end{align}
where $\theta_h$ is the mixing angle. Combining the Eq. \ref{metan}-\ref{etamix}, we get the masses of the $\eta_{1}^L$ and $\eta_{1}^H$ as
\begin{align}
    m_{\eta_1^L}^2 &= m_{\pi_1}^2 + \delta_S^{hyb}(1-\sec(2\theta_h))\\
    m_{\eta_1^H}^2 &= m_{\pi_1}^2 + \delta_S^{hyb}(1+\sec(2\theta_h)) \text{ .}
\end{align}
Note, since the hybrid isoscalars belong to a so-called homochiral multiplet (\cite{Giacosa:2017pos}) we do not expect them to mix significantly. Accordingly, the light isoscalar is expected to have a mass similar to (or lower than) the $\pi_1(1600)$, and the heavy one can be identified with the very recently discovered  resonance $\eta_1(1855)$.
\begin{table*}[ht]
    \centering
    {\renewcommand{\arraystretch}{1.5}
    \begin{tabular}{|c|c|c|c|c|c|}
    \hline
        Channel & \multicolumn{2}{c|}{Width (MeV)} & Channel & \multicolumn{2}{c|}{Width (MeV)}\\\cline{2-3}\cline{5-6}
        & Set$-1$ & Set$-2$ & & Set$-1$ & Set$-2$\\\hline
        $\Gamma_{K_1(1270)\pi}$ & $125\pm 42$ & $48\pm 25$ & $\Gamma_{\rho K}$ & $2.18\pm 0.56$ & $2.19\pm0.57$\\
        $\Gamma_{K_1(1400)\pi}$ & $103\pm 45$ & $98\pm 43$ & $\Gamma_{\omega K}$ & $0.82\pm 0.21$ & $0.82\pm0.21$\\
        $\Gamma_{h_1(1170)K}$ & $1.53\pm 0.28$ & $1.37\pm 0.24$ & $\Gamma_{\phi K}$ & $0.49\pm 0.12$ & $0.49\pm0.13$\\
        $\Gamma_{\eta K}$ & $0.29\pm 0.07$ & $0.29\pm0.07$ & $\Gamma_{K^*\pi}$ & $0.67\pm 0.17$ & $0.67\pm0.17$\\\
        $\Gamma_{\eta^\prime K}$ & $2.77\pm 0.62$ & $2.81\pm0.62$ & $\Gamma_{K^*\eta}$ & $0.30\pm 0.08$ & $0.30\pm0.08$\\
        $\Gamma_{\rho K^*}$ & $0.045\pm 0.016$ & $0.047\pm 0.016$ & $\Gamma_{\omega K^*}$ & $0.011\pm 0.004$ & $0.012\pm 0.004$ \\
        $\Gamma_{a_1 K}$ & $11.0\pm 2.32$ & $11.3\pm 2.35$ & $\Gamma_{b_1 K}$ & $64\pm 14$ & $3.11\pm 2.88$ \\\hline
        & & & $\Gamma_\text{tot}$ & $312\pm 97$ & $170\pm 65$\\\hline
    \end{tabular}}
    \caption{The partial widths and branching ratios of various decay channels and the total width for the hybrid kaon $K_1^{hyb}(1750)$. We have assumed the mass of the state to be $1761$ MeV \cite{Eshraim:2020ucw}.}
    \label{decwidkaon}
\end{table*}

Applying the recently reported data \cite{Ablikim:2022zze} to the analysis done in Ref. \cite{Eshraim:2020ucw}, we find that the observed mass of the $\eta_1(1855)$ can be explained in three possible ways:
\begin{enumerate}
    \item Scenario-1: The strangeness contribution to the hybrid $\eta_{1,S}$ state is the same as the ones to the conventional states, {\it i.e.,} $\delta_S^{hyb} = 0.151$ GeV$^2$ \cite{Eshraim:2020ucw,Parganlija:2012fy} (see, Table \ref{parscen}). This would imply that for the $\eta_1^H$ to be identified with the $\eta_1(1855)$, a large mixing angle ($\theta_h\sim 36.7^\circ$) would be needed, contrary to the homochiral nature of the states \cite{Giacosa:2017pos,Eshraim:2020ucw}. The total width of the $\eta_1^H$ would then be $607\pm159$ MeV ($249\pm 80$ MeV) for the parameter set-1 (set-2) (see, Table \ref{tabsum}). Both these value are significantly different from the experimentally measured value of $188\pm 18^{+3}_{-8}$ MeV \cite{Ablikim:2022zze}.
    \item Scenario-2: The strangeness contribution provides bulk of the extra mass of the $\eta_1(1855)$. In this scenario, the mixing angle vanishes ($\theta_h = 0$), thus $\delta_S^{hyb}=0.341$ GeV$^2$. The kaon turns out to have a mass of $1761$ MeV, while the mass of the light isoscalar amounts to $1661$ MeV. This case, however, leads to a total width of $259\pm 92$ MeV ($157\pm 68$ MeV) for the $\eta_1^H$, which are in agreement with the experimental observation at $1\sigma$ level.
    \item Scenario-3: The mass of the $\eta_1(1855)$ is a consequence of the combination of the above two scenarios, {\it i.e.} the strangeness contribution accounts for some part of the extra mass and a non-zero (but small) mixing angle provides for the rest. For our purpose, we take $\theta_h=15^\circ$, giving $\delta_S^{hyb}=0.317$ GeV$^2$. The masses of the kaon and the light isoscalar would fall between the estimates of the previous two cases - at $1754$ MeV and $1646$ MeV respectively. This leads to a total width of $411\pm 130$ MeV ($192\pm 80$ MeV) for the parameter set-1 (set-2).
\end{enumerate}

The partial widths of the various decays of the isoscalar hybrids corresponding to scenario-2 are given in Table \ref{decwideta}. The $\eta_1^L$ has four open decay channels: $a_1(1260)\pi$, $K^*K$, $\eta\eta'$, and the $\rho\rho$. Of these, the $a_1(1260)\pi$ channel is the dominant one. 
The total width of this isoscalar is $\sim 81$ MeV, which is small compared to its siblings. It should be noted that the $K_1(1270/1400)K$, $f_1^{(\prime)}\eta^{(\prime)}$ channels are forbidden at tree-level as they are sub-threshold. These channels could eventually contribute a significant amount to the width of the $\eta_1^L$, leading to a somewhat broader state if the finite width of the states is taken into account (see below).\par

On the other hand, $\eta_1^H$ isoscalar can decay into $K_1(1270)K$, $K^*K$, $\eta\eta'$, $f_1\eta$, $K^*K^*$, and the $\omega\phi$ states. The $K_1(1270)K$ is expected to be the dominant decay channel, followed by the $a_1\pi$ channel. With these decay channels, the width of $\eta_1(1855)$ is $259\pm 92$ MeV (parameter set-1), which is consistent with the experimental value at $1\sigma$ level.\par

However, the broad nature of the (dominant) $a_1$ and the presence of sub-threshold channels could eventually modify width for this state, (possibly) requiring a tweaking of the mixing angle. \par
It is furthermore interesting that the $K_1(1270)K$ is always the dominant channel for the decay of the heavy $\eta_1^H$ and the decay channel $a_1\pi$ is always dominant for $\eta_1^L$, independently of the three scenarios that we have discussed.\par

The decays of the hybrid isoscalars involve broad states and a large number of sub-threshold channels. To get a full picture of the decays of the hybrid isoscalars, it becomes necessary to perform a spectral integration over the final states. A detailed study in this direction will be attempted in the future.\par

\subsection{The hybrid kaon: \texorpdfstring{$K_1^{hyb}(1750)$}{}}
The mass of the kaon is given by \cite{Eshraim:2020ucw},
\begin{align}
    m_{K_1}^2 &= m_{\pi_1}^2 + \delta_s^{hyb}.
\end{align}
Based on the discussion present in the previous subsection, we take the mass of the kaon as $1761$ MeV (i.e. Scenario-3). As seen in the Table \ref{decwidkaon}, the kaons are expected to be broad states, similar to the isovectors.\par
The hybrid kaons have more decay channels available at the tree level compared to the $\pi_1(1600)$. One can expect the hybrid kaons to decay into $K_1(1270)\pi$, $K_1(1400)\pi$, $a_1K$, $b_1K$, $h_1(1170)K$, $\rho K$, $\omega K$, $\phi K$, $K^*\pi$, $K^*\eta$, $\eta K$, $\eta'K$, $\rho K^*$, and $\omega K^*$. Of these the axial-kaon channels are expected to be the most dominant channels. The estimated partial widths of these decay channels are given in Table \ref{decwidkaon}. Apart from the axial-kaon decay channels, we also expect the hybrid kaon to decay into the $b_1 K$, $a_1 K$, and the $\eta'K$ channels with significant widths. The $\eta K$ channel is nearly one order of magnitude smaller than the $\eta^\prime K$ channel. The vector-vector decay channels ($\rho K^*$ and $\omega K^*$) appear to be strongly suppressed.\par
In the axial-kaonic channels, since the decay thresholds are very close to the mass of the kaon, we expect the $K_1(1270)\pi$, $K_1(1400)\pi$ , $a_1K$, and the $b_1K$ thresholds to distort the line shape of the hybrid kaon significantly. This is true for other channels as well (except for the $K^*\pi$ and the $\eta K$ channels), but their partial widths are too small to cause any significant change.\par

Another interesting observation is that the partial width of the $K_1^{hyb}(1750)$ $\to h_1(1170)K$ decay is sensitive to the $h_1(1170)-h_1'(1415)$ mixing angle. In the calculation of the partial widths listed in Table \ref{decwidkaon} we have assumed a mixing angle of $\theta_{pv}= +25^\circ$ as derived in \cite{Shastry:2021asu} (for other possible values, see, e.g., Refs. \cite{Cheng:2011pb,Cheng:2017pcq}). However, if the mixing angle is reduced to $\theta_{pv}=0.6^\circ$, as reported by the BESIII collaboration \cite{BESIII:2018ede}, we obtain the partial width to be $15.5\pm 2.8$ MeV ($14.0\pm 2.5$ MeV) for the parameter Set-1 (Set-2). This would also increase the total width of the hybrid kaon to $326\pm 97$ MeV ($182\pm 64$ MeV). On the other hand, if the mixing angle is negative ($\theta_{pv}=-25^\circ$), then the partial width increases to $36\pm 6.5$ MeV ($33\pm 5.8$) MeV, and the total width to $346\pm 99$ MeV ($201\pm 64$ MeV).
We thus expect the kaon to have a total width of $300-400$ MeV.\par

Finally, it should be recalled that the hybrid kaon state can, in principle, mix with nearby vector states, such as the excited vector kaons $K^*(1410)$ and $K^*(1680)$ \cite{Zyla:2020ssz}. At present, the results of $K^*(1410)$ and $K^*(1680)$ fit quite well with being purely quark-antiquark states \cite{Piotrowska:2017rgt,Feng:2021igh} belonging to the nonet of radially excited and orbitally excited vector kaons, hence we neglect this mixing. Further, even though the orbitally excited $\bar{q}q$ and the hybrid kaonic states have the same possible decay channels, their partial widths are ``complementary" in the sense that, the dominant decay channels of the $K^*(1680)$ are suppressed for the $K_1^{hyb}$ (for instance, $K^\ast(1680)$ strongly decays into $K\pi$, $\rho K$, and the $K^*\pi$ channels, which -according to our results- are suppressed for the hybrid kaon). Moreover, because of the difference in their masses, some of the channels available for the decay of the hybrid kaon ({\it i.e.,} $a_1K$, $b_1K$) are sub-threshold for the vector kaon. These features can be used to detect the existence of the hybrid kaon and can be an important test for our model.\par
Nevertheless, the mixing of vector kaonic states belonging to distinct nonets should be eventually reconsidered when experimental data about this yet putative state will be available. 

\section{Summary and Conclusions}\label{secsum}
In this work, we have studied the decays of the $1^{-+}$ hybrid nonet. The experimental and lattice results for the $\pi_1(1600)$ were implemented, together with flavor constraints, in a single fit. As an outcome, both the mass and the decay widths $\pi_1(1600)$ could be re-determined with a better accuracy. 
For definiteness, a Lagrangian invariant under the flavor symmetry, parity reversal, and charge conjugation was utilized for describing the various decays. 

Next, we have estimated the partial widths for the various allowed decay channels and also the total widths of the remaining members of the hybrid nonet (the putative states $K_1^{hyb}(1750)$, $\eta_1^L$, $\eta_1^H$). We are able to identify the $\eta_1^H$ with the $\eta_1(1855)$, whereas the light isoscalar  $\eta_1^{L}$ turns out to be rather narrow.

The quantities estimated in this work can be used as guiding values for the ongoing experimental searches for the missing hybrid states $\eta_1^{L}$ and $K_1^{hyb} $\cite{Hamdi:2019dbr} and toward a better understanding of $\pi_1(1600)$ as well as the very recently discovered $\eta_1(1855)$ (for which an independent confirmation is needed). We expect the $1^{-+}$ kaonic and isoscalar hybrids to be observable in the $K3\pi$, $4\pi$ and $2\pi2K$ channels, respectively. Needless to say, the dsicovery and assesment of the lightest hybrid nonet would constitute an important step forward in low-energy QCD. 

A more complete description of the decays studied in this work can be obtained by expanding the formalism to the sub-threshold decay channels using spectral integration. This can lead to interesting results, as there are quite a few decay channels with thresholds slightly greater than the masses of the hybrids. Using spectral distribution functions that inherently take into account the thresholds ({\it e.g,} Ref. \cite{Giacosa:2021mbz}), one can study the contributions of these decay channels. This will be attempted in the future.\par

\section*{Acknowledgements}
F. G. and V. S. acknowledge financial support from the Polish National Science Centre (NCN) via the OPUS project 2019/ 33/B/ST2/00613. F.G. acknowledges also support from the NCN OPUS project no. 2018/29/B/ST2/02576. C.F. is supported by BMBF under grant number 05P21RGFP3.


\begin{thebibliography}{100}
\bibitem{Brambilla:2019esw}
N.~Brambilla, S.~Eidelman, C.~Hanhart, A.~Nefediev, C.~P.~Shen, C.~E.~Thomas, A.~Vairo and C.~Z.~Yuan,
``The $XYZ$ states: experimental and theoretical status and perspectives,''
Phys. Rept. \textbf{873}, 1-154 (2020)
doi:10.1016/j.physrep.2020.05.001
[arXiv:1907.07583 [hep-ex]].

\bibitem{Zyla:2020ssz}
P.~A.~Zyla \textit{et al.} [Particle Data Group],
``Review of Particle Physics,''
PTEP \textbf{2020}, no.8, 083C01 (2020)
doi:10.1093/ptep/ptaa104

\bibitem{Ablikim:2022zze}
M.~Ablikim, M.~N.~Achasov, P.~Adlarson, S.~Ahmed, M.~Albrecht, R.~Aliberti, A.~Amoroso, M.~R.~An, Q.~An and X.~H.~Bai, \textit{et al.}
``Observation of an isoscalar resonance with exotic $J^{PC}=1^{-+}$ quantum numbers in $J/\psi\rightarrow\gamma\eta\eta'$,''
[arXiv:2202.00621 [hep-ex]].

\bibitem{Chen:2022qpd}
H.~X.~Chen, N.~Su and S.~L.~Zhu,
``QCD axial anomaly enhances the $\eta \eta^\prime$ decay of the hybrid candidate $\eta_1(1855)$,''
[arXiv:2202.04918 [hep-ph]].

\bibitem{Qiu:2022ktc}
L.~Qiu and Q.~Zhao,
``Towards the establishment of the light $J^{P(C)}=1^{-(+)}$ hybrid nonet,''
[arXiv:2202.00904 [hep-ph]].

\bibitem{Dong:2022cuw}
X.~K.~Dong, Y.~H.~Lin and B.~S.~Zou,
``Interpretation of the $\eta_1(1855)$ as a $K\bar K_1(1400)+$ c.c. molecule,''
[arXiv:2202.00863 [hep-ph]].

\bibitem{E852:1998mbq}
G.~S.~Adams \textit{et al.} [E852],
``Observation of a new J(PC) = 1-+ exotic state in the reaction pi- p --\ensuremath{>} pi+ pi- pi- p at 18-GeV/c,''
Phys. Rev. Lett. \textbf{81}, 5760-5763 (1998)
doi:10.1103/PhysRevLett.81.5760

\bibitem{Meyer:2015eta}
C.~A.~Meyer and E.~S.~Swanson,
``Hybrid Mesons,''
Prog. Part. Nucl. Phys. \textbf{82} (2015), 21-58
doi:10.1016/j.ppnp.2015.03.001
[arXiv:1502.07276 [hep-ph]].

\bibitem{E862:2006cfp}
G.~S.~Adams \textit{et al.} [E862],
``Confirmation of a pi(1)0 Exotic Meson in the eta pi0 System,''
Phys. Lett. B \textbf{657} (2007), 27-31
doi:10.1016/j.physletb.2007.07.068
[arXiv:hep-ex/0612062 [hep-ex]].

\bibitem{OBELIX:2004oio}
P.~Salvini \textit{et al.} [OBELIX],
``anti-p p annihilation into four charged pions at rest and in flight,''
Eur. Phys. J. C \textbf{35} (2004), 21-33
doi:10.1140/epjc/s2004-01811-8

\bibitem{CrystalBarrel:1999reg}
A.~Abele \textit{et al.} [Crystal Barrel],
``Evidence for a pi eta P wave in anti-p p annihilations at rest into pi0 pi0 eta,''
Phys. Lett. B \textbf{446} (1999), 349-355
doi:10.1016/S0370-2693(98)01544-5

\bibitem{CrystalBarrel:1998cfz}
A.~Abele \textit{et al.} [Crystal Barrel],
``Exotic eta pi state in anti-p d annihilation at rest into pi- pi0 eta p(spectator),''
Phys. Lett. B \textbf{423} (1998), 175-184
doi:10.1016/S0370-2693(98)00123-3

\bibitem{Bass:2001zs}
S.~D.~Bass and E.~Marco,
``Final state interaction and a light mass 'exotic' resonance,''
Phys. Rev. D \textbf{65}, 057503 (2002)
doi:10.1103/PhysRevD.65.057503
[arXiv:hep-ph/0108189 [hep-ph]].

\bibitem{JPAC:2018zyd}
A.~Rodas \textit{et al.} [JPAC],
``Determination of the pole position of the lightest hybrid meson candidate,''
Phys. Rev. Lett. \textbf{122}, no.4, 042002 (2019)
doi:10.1103/PhysRevLett.122.042002
[arXiv:1810.04171 [hep-ph]].

\bibitem{Hamdi:2019dbr}
A.~Hamdi [GlueX],
``Search for exotic states in photoproduction at GlueX,''
J. Phys. Conf. Ser. \textbf{1667} (2020) no.1, 012012
doi:10.1088/1742-6596/1667/1/012012
[arXiv:1908.11786 [nucl-ex]].

\bibitem{COMPASS:2021ogp}
G.~D.~Alexeev \textit{et al.} [COMPASS],
``Exotic meson $\pi_1(1600)$ with $J^{PC} = 1^{-+}$ and its decay into $\rho(770)\pi$,''
Phys. Rev. D \textbf{105} (2022) no.1, 012005
doi:10.1103/PhysRevD.105.012005
[arXiv:2108.01744 [hep-ex]].

\bibitem{BESIII:2022qzu}
M.~Ablikim \textit{et al.} [BESIII],
``Partial wave analysis of $J/\psi\rightarrow\gamma\eta\eta'$,''
[arXiv:2202.00623 [hep-ex]].

\bibitem{Isgur:1985vy}
N.~Isgur, R.~Kokoski and J.~Paton,
``Gluonic Excitations of Mesons: Why They Are Missing and Where to Find Them,''
Phys. Rev. Lett. \textbf{54} (1985), 869
doi:10.1103/PhysRevLett.54.869

\bibitem{Page:1998gz}
P.~R.~Page, E.~S.~Swanson and A.~P.~Szczepaniak,
``Hybrid meson decay phenomenology,''
Phys. Rev. D \textbf{59} (1999), 034016
doi:10.1103/PhysRevD.59.034016
[arXiv:hep-ph/9808346 [hep-ph]].

\bibitem{Burns:2006wz}
T.~Burns and F.~E.~Close,
``Hybrid meson properties in Lattice QCD and Flux Tube Models,''
Phys. Rev. D \textbf{74} (2006), 034003
doi:10.1103/PhysRevD.74.034003
[arXiv:hep-ph/0604161 [hep-ph]].

\bibitem{Close:1994hc}
F.~E.~Close and P.~R.~Page,
``The Production and decay of hybrid mesons by flux tube breaking,''
Nucl. Phys. B \textbf{443} (1995), 233-254
doi:10.1016/0550-3213(95)00085-7
[arXiv:hep-ph/9411301 [hep-ph]].

\bibitem{Swanson:1997wy}
E.~S.~Swanson and A.~P.~Szczepaniak,
``Hybrid decays,''
Phys. Rev. D \textbf{56} (1997), 5692-5695
doi:10.1103/PhysRevD.56.5692
[arXiv:hep-ph/9704434 [hep-ph]].

\bibitem{Chanowitz:1982qj}
M.~S.~Chanowitz and S.~R.~Sharpe,
``Hybrids: Mixed States of Quarks and Gluons,''
Nucl. Phys. B \textbf{222} (1983), 211-244
[erratum: Nucl. Phys. B \textbf{228} (1983), 588-588]
doi:10.1016/0550-3213(83)90635-1

\bibitem{Chen:2010ic}
H.~X.~Chen, Z.~X.~Cai, P.~Z.~Huang and S.~L.~Zhu,
``The Decay Properties of the $1^{-+}$ Hybrid State,''
Phys. Rev. D \textbf{83} (2011), 014006
doi:10.1103/PhysRevD.83.014006
[arXiv:1010.3974 [hep-ph]].

\bibitem{Huang:2010dc}
P.~Z.~Huang, H.~X.~Chen and S.~L.~Zhu,
``The Strong Decay Patterns of the $1^{-+}$ Exotic Hybrid Mesons,''
Phys. Rev. D \textbf{83} (2011), 014021
doi:10.1103/PhysRevD.83.014021
[arXiv:1010.2293 [hep-ph]].

\bibitem{Qin:2011xq}
S.~x.~Qin, L.~Chang, Y.~x.~Liu, C.~D.~Roberts and D.~J.~Wilson,
``Investigation of rainbow-ladder truncation for excited and exotic mesons,''
Phys. Rev. C \textbf{85} (2012), 035202
doi:10.1103/PhysRevC.85.035202
[arXiv:1109.3459 [nucl-th]].

\bibitem{Burden:2002ps}
C.~J.~Burden and M.~A.~Pichowsky,
``J(PC) exotic mesons from the Bethe-Salpeter equation,''
Few Body Syst. \textbf{32} (2002), 119-126
doi:10.1007/s00601-002-0113-5
[arXiv:hep-ph/0206161 [hep-ph]].


\bibitem{Krassnigg:2009zh}
A.~Krassnigg,
``Survey of J=0,1 mesons in a Bethe-Salpeter approach,''
Phys. Rev. D \textbf{80} (2009), 114010
doi:10.1103/PhysRevD.80.114010
[arXiv:0909.4016 [hep-ph]].


\bibitem{Hilger:2015hka}
T.~Hilger, M.~Gomez-Rocha and A.~Krassnigg,
``Masses of $J^{PC} =1^{-+}$ exotic quarkonia in a Bethe-Salpeter-equation approach,''
Phys. Rev. D \textbf{91} (2015) no.11, 114004
doi:10.1103/PhysRevD.91.114004
[arXiv:1503.08697 [hep-ph]].

\bibitem{Williams:2015cvx}
R.~Williams, C.~S.~Fischer and W.~Heupel,
``Light mesons in QCD and unquenching effects from the 3PI effective action,''
Phys. Rev. D \textbf{93} (2016) no.3, 034026
doi:10.1103/PhysRevD.93.034026
[arXiv:1512.00455 [hep-ph]].

\bibitem{Xu:2018cor}
S.~S.~Xu, Z.~F.~Cui, L.~Chang, J.~Papavassiliou, C.~D.~Roberts and H.~S.~Zong,
``New perspective on hybrid mesons,''
Eur. Phys. J. A \textbf{55} (2019) no.7, 113
doi:10.1140/epja/i2019-12805-4
[arXiv:1805.06430 [nucl-th]].

\bibitem{Lan:2021wok}
J.~Lan \textit{et al.} [BLFQ],
``Light mesons with one dynamical gluon on the light front,''
Phys. Lett. B \textbf{825} (2022), 136890
doi:10.1016/j.physletb.2022.136890
[arXiv:2106.04954 [hep-ph]].

\bibitem{Kopf:2020yoa}
B.~Kopf, M.~Albrecht, H.~Koch, M.~K\"u\ss{}ner, J.~Pychy, X.~Qin and U.~Wiedner,
``Investigation of the lightest hybrid meson candidate with a coupled-channel analysis of ${{\bar{p}}p}$-, $\pi ^- p$- and ${\pi \pi }$-Data,''
Eur. Phys. J. C \textbf{81} (2021) no.12, 1056
doi:10.1140/epjc/s10052-021-09821-2
[arXiv:2008.11566 [hep-ph]].

\bibitem{Dudek:2010wm}
J.~J.~Dudek, R.~G.~Edwards, M.~J.~Peardon, D.~G.~Richards and C.~E.~Thomas,
``Toward the excited meson spectrum of dynamical QCD,''
Phys. Rev. D \textbf{82} (2010), 034508
doi:10.1103/PhysRevD.82.034508
[arXiv:1004.4930 [hep-ph]].

\bibitem{Dudek:2011bn}
J.~J.~Dudek,
``The lightest hybrid meson supermultiplet in QCD,''
Phys. Rev. D \textbf{84} (2011), 074023
doi:10.1103/PhysRevD.84.074023
[arXiv:1106.5515 [hep-ph]].

\bibitem{Dudek:2011tt}
J.~J.~Dudek, R.~G.~Edwards, B.~Joo, M.~J.~Peardon, D.~G.~Richards and C.~E.~Thomas,
``Isoscalar meson spectroscopy from lattice QCD,''
Phys. Rev. D \textbf{83} (2011), 111502
doi:10.1103/PhysRevD.83.111502
[arXiv:1102.4299 [hep-lat]].

\bibitem{Dudek:2013yja}
J.~J.~Dudek \textit{et al.} [Hadron Spectrum],
``Toward the excited isoscalar meson spectrum from lattice QCD,''
Phys. Rev. D \textbf{88} (2013) no.9, 094505
doi:10.1103/PhysRevD.88.094505
[arXiv:1309.2608 [hep-lat]].

\bibitem{Lacock:1996vy}
P.~Lacock \textit{et al.} [UKQCD],
``Orbitally excited and hybrid mesons from the lattice,''
Phys. Rev. D \textbf{54} (1996), 6997-7009
doi:10.1103/PhysRevD.54.6997
[arXiv:hep-lat/9605025 [hep-lat]].

\bibitem{Lacock:1996ny}
P.~Lacock \textit{et al.} [UKQCD],
``Hybrid mesons from quenched QCD,''
Phys. Lett. B \textbf{401} (1997), 308-312
doi:10.1016/S0370-2693(97)00384-5
[arXiv:hep-lat/9611011 [hep-lat]].

\bibitem{MILC:1997usn}
C.~W.~Bernard \textit{et al.} [MILC],
``Exotic mesons in quenched lattice QCD,''
Phys. Rev. D \textbf{56} (1997), 7039-7051
doi:10.1103/PhysRevD.56.7039
[arXiv:hep-lat/9707008 [hep-lat]].

\bibitem{McNeile:2006bz}
C.~McNeile \textit{et al.} [UKQCD],
``Decay width of light quark hybrid meson from the lattice,''
Phys. Rev. D \textbf{73} (2006), 074506
doi:10.1103/PhysRevD.73.074506
[arXiv:hep-lat/0603007 [hep-lat]].

\bibitem{Woss:2020ayi}
A.~J.~Woss \textit{et al.} [Hadron Spectrum],
``Decays of an exotic $1{-+}$ hybrid meson resonance in QCD,''
Phys. Rev. D \textbf{103}, no.5, 054502 (2021)
doi:10.1103/PhysRevD.103.054502
[arXiv:2009.10034 [hep-lat]].

\bibitem{Shastry:2021asu}
V.~Shastry, E.~Trotti and F.~Giacosa,
``Constraints imposed by the partial wave amplitudes on the decays of $J=1,2$ mesons,''
[arXiv:2107.13501 [hep-ph]].

\bibitem{Eshraim:2020ucw}
W.~I.~Eshraim, C.~S.~Fischer, F.~Giacosa and D.~Parganlija,
``Hybrid phenomenology in a chiral approach,''
Eur. Phys. J. Plus \textbf{135} (2020) no.12, 945
doi:10.1140/epjp/s13360-020-00900-z
[arXiv:2001.06106 [hep-ph]].

\bibitem{Amelino-Camelia:2010cem}
G.~Amelino-Camelia, F.~Archilli, D.~Babusci, D.~Badoni, G.~Bencivenni, J.~Bernabeu, R.~A.~Bertlmann, D.~R.~Boito, C.~Bini and C.~Bloise, \textit{et al.}
``Physics with the KLOE-2 experiment at the upgraded DA$\phi$NE,''
Eur. Phys. J. C \textbf{68}, 619-681 (2010)
doi:10.1140/epjc/s10052-010-1351-1
[arXiv:1003.3868 [hep-ex]].

\bibitem{LHCb:2013ged}
R.~Aaij \textit{et al.} [LHCb],
``Observation of $\bar{B}_{(s)} \to J/\psi f_1$(1285) Decays and Measurement of the $f_1$(1285) Mixing Angle,''
Phys. Rev. Lett. \textbf{112} (2014) no.9, 091802
doi:10.1103/PhysRevLett.112.091802
[arXiv:1310.2145 [hep-ex]].

\bibitem{Divotgey:2013jba}
F.~Divotgey, L.~Olbrich and F.~Giacosa,
``Phenomenology of axial-vector and pseudovector mesons: decays and mixing in the kaonic sector,''
Eur. Phys. J. A \textbf{49} (2013), 135
doi:10.1140/epja/i2013-13135-3
[arXiv:1306.1193 [hep-ph]].

\bibitem{Cheng:2011pb}
H.~Y.~Cheng,
``Revisiting Axial-Vector Meson Mixing,''
Phys. Lett. B \textbf{707} (2012), 116-120
doi:10.1016/j.physletb.2011.12.013
[arXiv:1110.2249 [hep-ph]].

\bibitem{COMPASS:2018uzl}
M.~Aghasyan \textit{et al.} [COMPASS],
``Light isovector resonances in $\pi^- p \to \pi^-\pi^-\pi^+ p$ at 190 GeV/${\it c}$,''
Phys. Rev. D \textbf{98} (2018) no.9, 092003
doi:10.1103/PhysRevD.98.092003
[arXiv:1802.05913 [hep-ex]].

\bibitem{E852:2004rfa}
M.~Lu \textit{et al.} [E852],
``Exotic meson decay to omega pi0 pi-,''
Phys. Rev. Lett. \textbf{94} (2005), 032002
doi:10.1103/PhysRevLett.94.032002
[arXiv:hep-ex/0405044 [hep-ex]].

\bibitem{E852:2001ikk}
E.~I.~Ivanov \textit{et al.} [E852],
``Observation of exotic meson production in the reaction pi- p ---\ensuremath{>} eta-prime pi- p at 18-GeV / c,''
Phys. Rev. Lett. \textbf{86} (2001), 3977-3980
doi:10.1103/PhysRevLett.86.3977
[arXiv:hep-ex/0101058 [hep-ex]].

\bibitem{E852:2004gpn}
J.~Kuhn \textit{et al.} [E852],
``Exotic meson production in the f(1)(1285) pi- system observed in the reaction pi- p ---\ensuremath{>} eta pi+ pi- pi- p at 18-GeV/c,''
Phys. Lett. B \textbf{595} (2004), 109-117
doi:10.1016/j.physletb.2004.05.032
[arXiv:hep-ex/0401004 [hep-ex]].

\bibitem{Baker:2003jh}
C.~A.~Baker, C.~J.~Batty, K.~Braune, D.~V.~Bugg, N.~Djaoshvili, W.~D\"unnweber, M.~A.~Faessler, F.~Meyer-Wildhagen, L.~Montanet and I.~Uman, \textit{et al.}
``Confirmation of a0(1450) and pi1(1600) in anti-p p --\ensuremath{>} omega pi+ pi- pi0 at rest,''
Phys. Lett. B \textbf{563} (2003), 140-149
doi:10.1016/S0370-2693(03)00643-9

\bibitem{Giacosa:2017pos}
F.~Giacosa, A.~Koenigstein and R.~D.~Pisarski,
``How the axial anomaly controls flavor mixing among mesons,''
Phys. Rev. D \textbf{97}, no.9, 091901 (2018)
doi:10.1103/PhysRevD.97.091901
[arXiv:1709.07454 [hep-ph]].

\bibitem{Parganlija:2012fy}
D.~Parganlija, P.~Kovacs, G.~Wolf, F.~Giacosa and D.~H.~Rischke,
``Meson vacuum phenomenology in a three-flavor linear sigma model with (axial-)vector mesons,''
Phys. Rev. D \textbf{87} (2013) no.1, 014011
doi:10.1103/PhysRevD.87.014011
[arXiv:1208.0585 [hep-ph]].

\bibitem{Cheng:2017pcq}
H.~Y.~Cheng and X.~W.~Kang,
``Branching fractions of semileptonic $D$ and $D_s$ decays from the covariant light-front quark model,''
Eur. Phys. J. C \textbf{77} (2017) no.9, 587
[erratum: Eur. Phys. J. C \textbf{77} (2017) no.12, 863]
doi:10.1140/epjc/s10052-017-5170-5
[arXiv:1707.02851 [hep-ph]].

\bibitem{BESIII:2018ede}
M.~Ablikim \textit{et al.} [BESIII],
``Observation of $h_1(1380)$ in the $J/\psi \to \eta^{\prime} K\bar K \pi$ decay,''
Phys. Rev. D \textbf{98} (2018) no.7, 072005
doi:10.1103/PhysRevD.98.072005
[arXiv:1804.05536 [hep-ex]].

\bibitem{Piotrowska:2017rgt}
M.~Piotrowska, C.~Reisinger and F.~Giacosa,
``Strong and radiative decays of excited vector mesons and predictions for a new $\phi(1930)$ resonance,''
Phys. Rev. D \textbf{96} (2017) no.5, 054033
doi:10.1103/PhysRevD.96.054033
[arXiv:1708.02593 [hep-ph]].

\bibitem{Feng:2021igh}
J.~C.~Feng, X.~W.~Kang, Q.~F.~L\"u and F.~S.~Zhang,
``Possible assignment of excited light S31 vector mesons,''
Phys. Rev. D \textbf{104} (2021) no.5, 054027
doi:10.1103/PhysRevD.104.054027
[arXiv:2104.01339 [hep-ph]].

\bibitem{Giacosa:2021mbz}
F.~Giacosa, A.~Okopi\'nska and V.~Shastry,
``A simple alternative to the relativistic Breit\textendash{}Wigner distribution,''
Eur. Phys. J. A \textbf{57} (2021) no.12, 336
doi:10.1140/epja/s10050-021-00641-2
[arXiv:2106.03749 [hep-ph]].
\end{thebibliography}
\end{document}